\documentclass[journal,12pt,draftclsnofoot,onecolumn]{IEEEtran}
\usepackage{graphicx}
\usepackage{times}
\usepackage[centertags]{amsmath}
\usepackage{amsfonts}
\usepackage{amssymb}
\usepackage{amsthm}
\usepackage{units}

\usepackage[tight,footnotesize]{subfigure}
\newtheorem{prop}{Proposition}
\newtheorem{exap}{Example}
\newtheorem{rem}{Remark}

\begin{document}

\title{Wyner-Ziv Coding Based on Multidimensional Nested Lattices}
%
%
%

\author{Cong Ling, Su Gao, and
Jean-Claude Belfiore
\thanks{The material in this paper was presented in part at the
International Conference on Communications, Dresden, Germany, June
2009. }
\thanks{Cong Ling and Su Gao are with the Department of Electrical and Electronic Engineering, Imperial College London, London SW7
2AZ, United Kingdom (e-mail: cling@ieee.org,
su.gao06@imperial.ac.uk).}
\thanks{Jean-Claude Belfiore is with the Department of Communications and Electronics, Telecom ParisTech, Paris, France (e-mail: belfiore@telecom-paristech.fr).}}



\maketitle

\begin{abstract}
Distributed source coding (DSC) addresses the compression of
correlated sources without communication links among them. This
paper is concerned with the Wyner-Ziv problem: coding of an
information source with side information available only at the
decoder in the form of a noisy version of the source. Both the
theoretical analysis and code design are addressed in the framework
of multi-dimensional nested lattice coding (NLC). For theoretical
analysis, accurate computation of the rate-distortion function is
given under the high-resolution assumption, and a new upper bound
using the derivative of the theta series is derived. For practical
code design, several techniques with low complexity are proposed.
Compared to the existing Slepian-Wolf coded nested quantization
(SWC-NQ) for Wyner-Ziv coding based on one or two-dimensional
lattices, our proposed multi-dimensional NLC can offer better
performance at arguably lower complexity, since it does not require
the second stage of Slepian-Wolf coding.

\end{abstract}

\begin{IEEEkeywords}
Distributed source coding, multi-dimensional lattices, nested lattices, rate-distortion function,
Wyner-Ziv coding.
\end{IEEEkeywords}

%
\IEEEpeerreviewmaketitle

\section{Introduction}
%
%
%
%
\IEEEPARstart{I}{n} wireless sensor networks, energy is a major
constraint to individual node performance. Compressing data prior to
transmission and minimizing inter-node communications can
effectively reduce energy consumption. The problem is well-known as
multi-terminal source coding or distributed source coding
(DSC)~\cite {xiong04}, where one or more of the sensor nodes (or
terminals, information sources) compress data separately (i.e.
without communication with each other) before transmission. The
joint decoder at the central node recovers the transmitted data
either losslessly or up to a prescribed distortion. For the lossless
case, Slepian and Wolf \cite{slepian73} proposed the theoretical
framework. A common scenario for the lossy case is the Wyner-Ziv
problem, which considers encoding a source where the decoder has
access to some side information~\cite {wyner76}.

Code designs for Wyner-Ziv coding have emerged in recent years. A
structured binning scheme using (random) nested lattice codes (NLC)
was introduced~\cite{Zamir98,zamir02}, which can asymptotically
achieve the Wyner-Ziv limit~\cite{wyner76} as the dimension of the
lattice approaches infinity. The first practical code design based
on this idea was done in~\cite{servetto00}, where preliminary
analysis for particular nested lattices was presented. A general
distortion analysis for $n$-dimensional lattices was given by Liu et
al.~\cite{liu06}. In particular, a lower bound on the distortion was
derived, which, for given $n$, exhibits an increasing gap to the
Wyner-Ziv limit as the rate grows. For this reason, \cite{liu06}
focused on $1$- and $2$-dimensional lattices and used a second stage
of coding, namely, Slepian-Wolf coding, to further exploit the
correlation between quantized data. This technique was termed as
``Slepian-Wolf coded nested quantization" (SWC-NQ) in~\cite{liu06}.
The Slepian-Wolf coding in \cite{liu06} relied on channel
capacity-achieving codes such as low-density parity check (LDPC)
codes, resulting in considerable implementation complexity.

In this paper, we take a new approach by directly using
multi-dimensional NLC, which is conceptually simpler. It was known
that the distortion performance improves with the lattice
dimension~\cite{zamir02,liu06}. Our work is inspired by this result,
and we aim to develop less complicated practical codes for the
Wyner-Ziv problem. The contributions of this paper are two-fold. On
one hand, we complement the theoretical analysis of the
rate-distortion function in~\cite{liu06}. Specifically, we develop a
technique to compute the distortion, which is accurate under the
high-resolution assumption, and a new upper bound based on the
derivative of the theta series. It is worth mentioning that, in
practice, an upper bound is a safer guideline of system design than
a lower bound. On the other hand, we implement NLC based on a
variety of techniques to obtain sublattices, and show that our
implementation gains over the conventional one and two-dimensional
schemes. These techniques include clean similar sublattices of
dimensions greater than two~\cite{conway99}, the random ensemble of
nested lattices in ~\cite{erez04}, and scaling and rotation of a
lattice to obtain its sublattices. Recently, a similar idea using
the scaling-rotating technique to obtain nested lattices was
proposed independently in \cite{fiore10} for the synchronous
multiple-access channel.

The obtained distortion performance is close to the Wyner-Ziv limit,
especially at rates $<10$ bits/sample. These simulation results are
consistent with theoretical analysis which predicts improved
performance as dimension increases, although the law of diminishing
returns will apply in high dimensions. Meanwhile, the complexity of
our proposed scheme is arguably lower than that of SWC-NQ, as we do
not use turbo or LDPC codes \cite{sartipi08} \cite{yang09}.

\section{Preliminaries}

Consider Wyner-Ziv coding in
the two-source case. Let $\{(X_i,Y_i) \}^\infty_{i=1} $ be a
sequence of independent and identically distributed (i.i.d.)
drawings of a pair of correlated random variables $X$ and $Y$, and
let $d(X_i,\widehat{X}_i)$ denote a single-letter distortion
measure between the source $X_i$ and its reconstructed version
$\widehat{X}_i$ at the decoder. Wyner-Ziv coding~\cite{wyner76} asks the question of how many bits are
needed to encode source $X$ under the constraint that the average
distortion $E[d(X_i,\widehat{X}_i)]$ is not greater than a given
target distortion $D$, assuming the side information $Y$ is
available at the decoder but not at the encoder.

We specifically consider the \emph{quadratic Gaussian} case of the
correlation model $X=Y+Z$, where $Z$ is the Gaussian noise with
distribution $Z\sim N(0,\sigma_Z^2)$, and $Y$ is independent of $Z$.
Mean-square error (MSE) is used for distortion measure. The
quadratic Gaussian case is special since it has no rate loss,
namely, the rate-distortion function is given by~\cite{liu06}
\begin{eqnarray}\label{EQ:WynerZivLimit}
D_{WZ}(R)=\sigma_Z^2 2^{-2R}
\end{eqnarray}
which will be referred to as the `Wyner-Ziv limit'.

\subsection{Lattices, nested lattices and similar sublattices}

For a set of $n$ linearly independent basis vectors
$\{\mathbf{m}_1,\mathbf{m}_2,\ldots,\mathbf{m}_n\}$, an
$n$-dimensional lattice $\Lambda$ is composed of all integral
combinations of the basis vectors:
\begin{equation}\label{eq:lambda}
\Lambda=\{\mathbf{l}=\mathbf{M} \cdot
\mathbf{i}:\mathbf{i}\in\mathbb{Z}^n\}
\end{equation}
where $\mathbf{M}=[\mathbf{m}_1,\mathbf{m}_2,\ldots,
\mathbf{m}_n]$ is the generator matrix and
$\mathbb{Z}=\{0,\pm1,\pm2,\ldots\}$ is the set of integers. For
a vector $\mathbf{x}=[x_1,x_2,\ldots,x_n]^T$, the
nearest-neighbor quantizer associated with $\Lambda$ is
$Q_{\Lambda}(\mathbf{x})=\arg\min_{\mathbf{l}\in\Lambda}||\mathbf{x}-\mathbf{l}||$.
The basic Voronoi cell of $\Lambda$, defined by
$\mathcal{V}=\{\mathbf{x}:Q_{\Lambda}(\mathbf{x})=\mathbf{0}\}$,
specifies the nearest-neighbor decoding region. Important
quantities for $\mathcal{V}$ include the cell volume
$V=\int_{\mathcal{V}}d\mathbf{x}$, the second moment
$\sigma^2=\frac{1}{nV}\int_{\mathcal{V}}||\mathbf{x}||^2d\mathbf{x}$
and the normalized second moment
$G(\Lambda)=\sigma^2/V^{\frac{2}{n}}$. The minimum of $G(\Lambda)$
of all the $n$-dimensional lattices is denoted as $G_n$.
From~\cite{conway1998book}, $G_n\ge1/2\pi e,\forall n$ and
$\lim_{n \to \infty}G_n=1/2\pi e$.

Let $\Lambda_F$ be a \emph{fine lattice} with a generator matrix
$\mathbf{M}_F$. Similarly, let $\Lambda_C$ be a \emph{coarse
lattice} with a generator matrix $\mathbf{M}_C$. A pair of
$n$-dimensional lattices $(\Lambda_F,\Lambda_C)$ is \emph{nested} in
the sense of $\Lambda_C\subset\Lambda_F$, if
$\mathbf{M}_C=\mathbf{M}_F\cdot\mathbf{P}$, where $\mathbf{P}$ is an
$n\times n$ integer matrix with determinant greater than one (in
absolute value). We define the \emph{nesting ratio} $N=V_C/V_F$,
where $V_F$ and $V_C$ are the cell volumes of the  fine and coarse
lattice, respectively.

There is one special case where $\Lambda_C$ is \emph{geometrically
similar} to $\Lambda_F$, which means that $\Lambda_C$ can be
obtained from $\Lambda_F$ by applying a similarity transform~\cite{conway99}
including a rotation, change of scale and possibly a reflection.
$\Lambda_C$ is \emph{strictly similar} to $\Lambda_F$ when
reflection is not used. We also refer to $\Lambda_C$ as a
\emph{similar sublattice} to $\Lambda_F$.

\subsection{Encoding and Decoding Scheme}

Throughout the paper, we follow the \emph{high-resolution
assumption} in~\cite{liu06}, which means $V_F$ is sufficiently small
such that the probability density function (pdf) of $Z$ is
approximately constant over each Voronoi cell of the fine lattice
$\Lambda_F$.

The encoding and decoding scheme is the same as in~\cite{liu06},
which is simplified from~\cite{zamir02} under the high-resolution
assumption. The scheme is described as follows:
\begin{itemize}
\item The encoder quantizes $\mathbf{x}$ to
$\mathbf{x}_{Q_{\Lambda_F}}=Q_{\Lambda_F}(\mathbf{x})$, compute the
coset leader
$\mathbf{s}=\mathbf{x}_{Q_{\Lambda_F}}-Q_{\Lambda_C}(\mathbf{x}_{Q_{\Lambda_F}})$,
and transmits the index corresponding to $\mathbf{s}$.
\item The decoder receives $\mathbf{s}$ and reconstructs
$\mathbf{x}$ as
$\widehat{\mathbf{x}}=\mathbf{s}+Q_{\Lambda_C}(\mathbf{y}-\mathbf{s})$.
\end{itemize}

Under this formulation, the rate per dimension is given by
$R=\frac{1}{n}\log_2 (\frac{V_C}{V_F})$. In our practical design, we
also use minimum MSE estimation as described in~\cite{liu06}.

\subsection{Theta series and even unimodular lattices}

The development of this paper is heavily based on the \emph{theta
series}. Given a lattice $\Lambda$, the {theta
series}~\cite{conway1998book} is defined as
\begin{eqnarray}
\Theta_{\Lambda}(q)=\sum_{\mathbf{x} \in \Lambda}
q^{\|\mathbf{x}\|^2}
\end{eqnarray}
where $q= e^{j\pi \tau}$ ($\Im(\tau)>0$). Using the change of
variable $j\tau\rightarrow \tau$ ($\tau>0$ real), it can
alternatively be expressed as
\begin{eqnarray}
\Theta_{\Lambda}(\tau)=\sum_{\mathbf{x} \in \Lambda} e^{-\pi \tau\|\mathbf{x}\|^2}.
\end{eqnarray}

Theta series of some standard lattices are given
in~\cite{conway1998book}. In general, theta series are not easy to
evaluate, and their derivatives are even more complex to calculate.
Fortunately, there is a class of lattices for which the computation
of such functions is tractable, the \emph{even unimodular lattices}.
A detailed description of such lattices can be found in
\cite{OSB11}.

A lattice $\Lambda$ is unimodular \cite{conway1998book} if
$\Lambda$ is integral, i.e., $\mathbf{M^T M}$ is an integer matrix, and $|\det\Lambda|=1$ or equavlently
$\Lambda$ is equal to its dual $\Lambda^*$, i.e.,
$\Lambda=\Lambda^*$. If $\Lambda$ is integral,
then $\|\mathbf{l}\|^2$ is necessarily an integer for
all $\mathbf{l}\in\Lambda$. Further, if $\|\mathbf{l}\|^2$ is an
even integer for all $\mathbf{l}\in\Lambda$, then $\Lambda$ is
called an even unimodular lattice. Many exceptional lattices such as the Gosset lattice $E_{8}$
or the Leech lattice $\Lambda_{24}$ are even unimodular.

We introduce now the three Jacobi theta functions
\cite{conway1998book},
\[
\begin{cases}
\vartheta_{2}\left(q\right) & =\sum_{n=-\infty}^{+\infty}q^{\left(n+\frac{1}{2}\right)^{2}}\\
\vartheta_{3}\left(q\right) & =\sum_{n=-\infty}^{+\infty}q^{n^{2}}\\
\vartheta_{4}\left(q\right) & =\sum_{n=-\infty}^{+\infty}(-1)^{n}q^{n^{2}}
\end{cases}.
\]

Now, consider the Eisenstein series,
\begin{equation}
E_{2k}\left(q\right)=1+\frac{2}{\zeta\left(1-2k\right)}\sum_{m=1}^{+\infty}\frac{m^{2k-1}q^{m}}{1-q^{m}}\label{eq:Eisenstein}
\end{equation}
where $\zeta\left(s\right)$ is the Riemann zeta function,
\[
\zeta(s)=\sum_{k=1}^{+\infty}\frac{1}{k^{s}}.
\]
We can relate $E_{4}\left(q\right)$ and the Jacobi theta functions
through the relation,
\[
E_{4}\left(q^{2}\right)=\frac{1}{2}\left(\vartheta_{2}^{8}\left(q\right)+\vartheta_{3}^{8}\left(q\right)+\vartheta_{4}^{8}\left(q\right)\right).
\]
Another fundamental series is the so-called modular discriminant defined
as,
\begin{equation}
\Delta\left(q\right)=\frac{1}{12^{3}}\left(E_{4}^{3}\left(q\right)-E_{6}^{2}\left(q\right)\right)\label{eq:discriminant}
\end{equation}
which is also related to the three Jacobi theta functions through
the identity
\[
\Delta\left(q^{2}\right)=\frac{1}{256}\vartheta_{2}^{8}\left(q\right)\vartheta_{3}^{8}\left(q\right)\vartheta_{4}^{8}\left(q\right).
\]

Remarkably, theta series of all even unimodular lattices can be expressed
as polynomials in the two variables $E_{4}$ and $\Delta$:
\begin{prop} [\cite{Ebeling}]
If $\Lambda$ is an even unimodular lattice of dimension $n$ then,
\begin{enumerate}
\item $n$ is a multiple of $8$, $n=24m+8k$ with $k=0,1,2$ and $m$ being
any non-negative integer;
\item its theta series is given by
\begin{equation}
\Theta_{\Lambda}\left(\tau\right)=E_{4}^{3m+k}\left(e^{-2\pi
\tau}\right)+\sum_{j=1}^{m}b_{j}E_{4}^{3(m-j)+k}\left(e^{-2\pi
\tau}\right)\Delta^{j}\left(e^{-2\pi \tau}\right),\quad
b_{j}\in\mathbb{Q}.\label{eq:unimod-theta}
\end{equation}

\end{enumerate}
\end{prop}

\section{New Analysis of Rate-Distortion Function}

Knowing the suitability of lattices to this problem through
theoretical analysis is essential for the code design. Our starting
point is the rate-distortion function of nested-lattice Wyner-Ziv
coding derived in~\cite{liu06}. It was shown there that the
distortion $D$ is comprised of a `source-coding component' $D_S$ and
a `channel-coding component' $D_C$. More precisely, the distortion
per dimension for the Wyner-Ziv coding at high resolution is given
by~\cite{liu06}
\begin{eqnarray}\label{EQ:LiuDistorFun}
D_n &=& \min_{V_C>0} \left\{D_S+D_C\right\}
\nonumber\\
&=& \min_{V_C>0} \left\{G(\Lambda_F)V_F^{\frac{2}{n}}+\frac{1}{n}E_{\mathbf{Z}}[\|Q_{\Lambda_C}(\mathbf{Z})\|^2]\right\}
\nonumber\\
&=& \min_{V_C>0} \left\{G(\Lambda_F)V_C^{\frac{2}{n}} 2^{-2R}
+\frac{1}{n}\sum_{\mathbf{l}_C\in
\Lambda_C}\|\mathbf{l}_C\|^2 P(\mathbf{z}\in
\mathcal{V}_C(\mathbf{l}_C))\right\}
\end{eqnarray}
where $D_S$ and $D_C$ are implicitly defined, and
$\mathcal{V}_C(\mathbf{l}_C)$ is the Voronoi cell associated with
the lattice point $\mathbf{l}_C\in \Lambda_C$.

There is a standard way to handle $D_S$: choosing a fine lattice
that is good for quantization, i.e., a lattice with $G(\Lambda_F)$
as small as possible. The second component $D_C$ is not so easy to
handle, though. This is the main problem we want to solve in this
paper.

A lower bound on $D_n$ was given in~\cite{liu06}, where the Voronoi
region $\mathcal{V}_C(\mathbf{l}_C)$ is replaced with the packing
sphere $\mathcal{S}$~\cite{conway1998book}. It was pointed out
in~\cite{liu06} that this lower bound is asymptotically tight since
the shape of $\mathcal{V}_C(\mathbf{l}_C)$ will approach a sphere as
$n$ increases for lattices with the best rate-distortion
performance. However, for finite $n$, the Voronoi regions may not be
close to a sphere. We observe that for these lattices, the lower
bound results in a non-negligible gap.

\subsection{Accurate Calculation Under High-Resolution Assumption}

We present an accurate calculation of the rate-distortion, based on
the high-resolution assumption.

\newtheorem{theorem}{Theorem}
\begin{theorem}[Accurate Calculation Under High-Resolution Assumption]\label{theorem:DcAccurate}
Let $\mathbf{l}_{F} \in \Lambda_F$ be a fine lattice point and let
\begin{eqnarray}\label{EQ:normpdf}
f_\mathbf{Z}(\mathbf{z})=\frac{1}{(\sqrt{2\pi}\sigma_Z)^n}\exp\left(-\frac{\|\mathbf{z}\|^2}{2\sigma_Z^2}\right)
\end{eqnarray}
be the pdf of a vector $\mathbf{z}$ whose elements $z_i$'s are
i.i.d. Gaussian with variance $\sigma_Z$. Then, under the
high-resolution assumption, the channel coding component $D_C$ of
the distortion function for nested-lattice Wyner-Ziv coding is given
by
\begin{eqnarray}\label{EQ:LiuDcHR}
D_C {=} \frac{V_F}{n}\sum_{\mathbf{l}_C\in
\Lambda_C}\|\mathbf{l}_C\|^2
\sum_{\mathbf{l}_{F} \in \mathcal{V}_C}f_\mathbf{Z}\left(\|\mathbf{l}_{F}+\mathbf{l}_C\|\right).
\end{eqnarray}

\end{theorem}

\begin{proof}
Based on the high-resolution assumption~\cite{liu06}, the pdf
$f_\mathbf{Z}(\mathbf{z})$ is uniform over each Voronoi cell of the
fine lattice. Thus $P(\mathbf{z}\in \mathcal{V}_C(\mathbf{l}_C))$ in
(\ref{EQ:LiuDistorFun}) can be approximated by
\begin{equation}
P\left(\mathbf{z}\in \mathcal{V}_C(\mathbf{l}_C)\right){=} \sum_{\mathbf{l}_{F} \in \mathcal{V}_C} f_{\mathbf{Z}}(\|\mathbf{l}_{F}+\mathbf{l}_C\|)V_F.
\end{equation}
Then the theorem follows.
\end{proof}

To compute (\ref{EQ:LiuDcHR}), we need to enumeration all lattice points $\mathbf{l}_{C} \in \Lambda_{C}$, and those points $\mathbf{l}_{F} \in \mathcal{V}_{C}$. The latter may be done by checking all points $\mathbf{l}_{F}$ within the covering radius of $\Lambda_{C}$:
$\mathbf{l}_{F} \in \mathcal{V}_{C}$ if and only if $\mathbf{l}_{F}$ is decoded to $0$ in $\Lambda_{C}$.

\begin{figure}[htbp]
  \centering
  \includegraphics{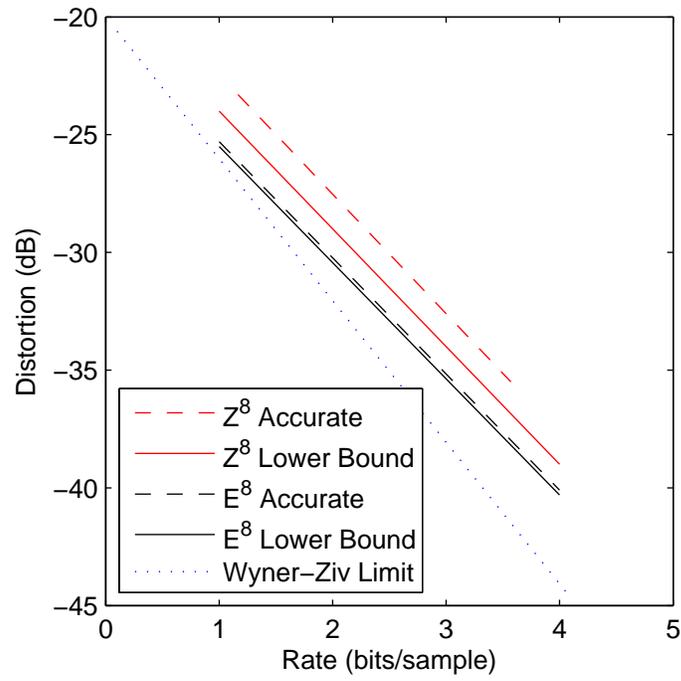}
  \caption{Comparison of proposed accurate calculation and the lower bound in~\cite{liu06}
  for fine lattices $\mathbb{Z}^8$ and $E^8$. $Y\sim N(0,1)$ and $Z\sim N(0,0.01)$.}
  \label{FIG:DvsRCompareAccurateAndLowerbound}
\end{figure}

In Fig.~\ref{FIG:DvsRCompareAccurateAndLowerbound}, we give the
rate-distortion performance of fine lattice $\mathbb{Z}^8$ and
$E_8$. The sublattice is generated from the expand-rotate method
described in the next Section. We also give the corresponding lower
bound in~\cite{liu06} for comparison. From
Fig.~\ref{FIG:DvsRCompareAccurateAndLowerbound} we can see in
dimension eight, the lower bound in~\cite{liu06} is tight for the
best quantization lattice $E_8$ in this dimension, but has a gap to
our theoretical calculation for the $\mathbb{Z}^8$ lattice. We find
this is also true for other dimensions.

\subsection{An Upper Bound}

Albeit accurate under the high-resolution assumption, the method in
the preceding subsection unfortunately does not lend much insight
into the selection criterion of a good coarse lattice. Let us
reexamine the $D_C$ component in (\ref{EQ:LiuDistorFun})
\begin{eqnarray} \label{eq:DC}
D_C&=&\frac{1}{n}\sum_{\mathbf{l}_C \in \Lambda_C} \|\mathbf{l}_C\|^2
\int_{\mathcal{V}_C} \frac{1}{(\sqrt{2\pi}\sigma_Z)^n}\exp\left(-\frac{\|\mathbf{l}_C+\mathbf{z}\|^2}{2\sigma_Z^2}\right)d\mathbf{z},
\end{eqnarray}
which should be minimized. Of course, this problem itself is not tractable due to the integration over the basic Voronoi region.

To deal with this problem, the common approach is to choose a coarse
lattice to maximize $P(\mathbf{z} \in \mathcal{V}_{C})$, which is
reduced to the standard approach to channel coding. However, this
standard approach does not quite capture the essence of Wyner-Ziv
coding, since the metric in (\ref{eq:DC}) is clearly different.
Alternatively, one could derive either the upper or lower bound by
using the covering or packing radius, the latter of which has
already been done by Liu et al. \cite{liu06}. Again, in addition to
the computational complexity, it is not insightful.

Here, we propose a new upper bound using the the derivative of the
theta series.

\begin{theorem} [Upper Bound on Distortion] \label{theorem:bound}

The component $D_C$ is bounded by
\begin{eqnarray} \label{EQ:UpperTheta}
D_C \leq -\frac{1}{2\pi n} \Theta_{\Lambda_C}^{'}\left(\frac{1}{8\pi \sigma_Z^2}\right)
\end{eqnarray}
where $\Theta_{\Lambda}^{'}$ denotes the derivative of the theta series (with respect to $\tau$):
\begin{eqnarray}
\Theta_{\Lambda}^{'}(\tau)=-\sum_{\mathbf{l} \in \Lambda} \pi \|\mathbf{l}\|^2e^{-\pi
\tau\|\mathbf{l}\|^2}.
\end{eqnarray}

\end{theorem}

\begin{proof}
The idea is to upper-bound the integral over ${\mathcal{V}_C}$ by
that over the half plane whose boundary is equally far from lattice
points $0$ and $\mathbf{l}_C$, similar to the technique widely used
in the union bound on the decoding error probability. That is,
$P(\mathbf{z}\in \mathcal{V}_C(\mathbf{l}_C)) \leq P(\|\mathbf{z}\|
> \|\mathbf{l}_C\|/2) = Q\left(\frac{\|\mathbf{l}_C\|}{2\sigma_Z}\right)$,
where
$Q(x)=\int_x^{\infty}\frac{1}{\sqrt{2\pi}}\exp\left(-\frac{t^2}{2}\right)dt$
for $x>0$ is the Gauss Q-function. Then, we have
\begin{eqnarray} \label{eq:upperbound}
D_C&\leq &\frac{1}{n}\sum_{\mathbf{l}_C \in \Lambda_C} \|\mathbf{l}_C\|^2 Q\left(\frac{\|\mathbf{l}_C\|}{2\sigma_Z}\right)
\nonumber\\
&\leq &\frac{1}{2n}\sum_{\mathbf{l}_C \in \Lambda_C} \|\mathbf{l}_C\|^2 \exp\left(-\frac{\|\mathbf{l}_C\|^2}{8\sigma_Z^2}\right),
\end{eqnarray}
where the second inequality follows from the fact $Q(x) \leq
\frac{1}{2}\exp(-x^2/2)$ for $x>0$.

Now we may recognize the right-hand side of (\ref{eq:upperbound}) as the derivative of the theta series.
Accordingly, the distortion can be expressed in the desired form.
\end{proof}

\begin{rem}
The bound may be slightly improved by applying the alternative
expression of the Q-function
$Q(x)=\frac{1}{\pi}\int_0^{\pi/2}\exp\left(-\frac{x^2}{2\sin^2\theta}\right)d\theta$
for $x>0$.
\end{rem}

Now the problem reduces to that of finding a coarse lattice which
maximizes the derivative of the theta series (note that its
derivative is negative). We formalize this new criterion as the
following Proposition:

\begin{prop} [Design Criterion of the Coarse Lattice]

A coarse lattice is good for Wyner-Ziv coding if the derivative of
its theta series is large.

\end{prop}

\begin{exap}

Table \ref{TAB:SeriesANDDeriv} shows the theta series and associated
derivatives of some well-known lattices.
\end{exap}

\begin{table}[t]
\centering \caption{Theta series and derivatives of well-known
lattices.} \label{TAB:SeriesANDDeriv}
\begin{tabular}{|c|c|c|}
\hline
   & Theta series & Derivative \\
\hline
$\mathbb{Z}^n$ & $\vartheta_3^n$ & $n\vartheta_3^{n-1}\vartheta_3^{'}$ \\
\hline $D_n$ & $\frac{1}{2}(\vartheta_3^n+\vartheta_4^n)$ &
$\frac{1}{2}(n\vartheta_3^{n-1}\vartheta_3^{'}+n\vartheta_4^{n-1}\vartheta_4^{'})$ \\
\hline $D_n^{\ast}$ & $\vartheta_2^n+\vartheta_3^n$ &
$n\vartheta_2^{n-1}\vartheta_2^{'}+n\vartheta_3^{n-1}\vartheta_3^{'}$ \\
\hline $E_8$ &
$\frac{1}{2}(\vartheta_2^8+\vartheta_3^8+\vartheta_4^8)$ &
$\frac{1}{2}(8\vartheta_2^{7}\vartheta_2^{'}+8\vartheta_3^{7}\vartheta_3^{'}+8\vartheta_4^{7}\vartheta_4^{'})$ \\
\hline
\end{tabular}

\end{table}

\subsection{Even Unimodular Lattices}

As the theta series of even unimodular lattices are polynomial in
$E_{4}$ and $\Delta$, it is enough to find the derivatives of these
two series. The Ramanujan system \cite{Ramanujan} gives the answer.
It expresses the derivatives of $E_{2}$, $E_{4}$ and $E_{6}$ as
functions of themselves:
\[
\begin{cases}
q\frac{dE_{2}(q)}{dq} & =\frac{1}{12}\left(E_{2}^{2}(q)-E_{4}(q)\right)\\
q\frac{dE_{4}(q)}{dq} & =\frac{1}{3}\left(E_{2}(q)E_{4}(q)-E_{6}(q)\right)\\
q\frac{dE_{6}(q)}{dq} & =\frac{1}{2}\left(E_{2}(q)E_{6}(q)-E_{4}^{2}(q)\right)
\end{cases}.
\]
As we are interested in the functions $E_{k}\left(e^{-2\pi
\tau}\right)$ for $k=4,6$, we get
\begin{eqnarray}
\frac{dE_{4}\left(e^{-2\pi \tau}\right)}{d\tau} & = & -\frac{2\pi}{3}\left(E_{2}\left(e^{-2\pi \tau}\right)E_{4}\left(e^{-2\pi \tau}\right)-E_{6}\left(e^{-2\pi \tau}\right)\right)\label{eq:E4-deriv}\\
\frac{dE_{6}\left(e^{-2\pi \tau}\right)}{d\tau} & = &
-\pi\left(E_{2}\left(e^{-2\pi \tau}\right)E_{6}\left(e^{-2\pi
\tau}\right)-E_{4}^{2}\left(e^{-2\pi
\tau}\right)\right)\label{eq:E6-deriv}
\end{eqnarray}
 and so, combining (\ref{eq:discriminant}), (\ref{eq:E4-deriv})
and (\ref{eq:E6-deriv}), we obtain the derivative of the modular
discriminant with respect to $\tau$:
\begin{equation}
\frac{d\Delta\left(e^{-2\pi \tau}\right)}{d\tau}=-2\pi
E_{2}\left(e^{-2\pi \tau}\right)\Delta\left(e^{-2\pi
\tau}\right)=-\frac{\pi}{6\times12^{2}}E_{2}\left(e^{-2\pi
\tau}\right)\left(E_{4}^{3}\left(e^{-2\pi
\tau}\right)-E_{6}^{2}\left(e^{-2\pi
\tau}\right)\right).\label{eq:Delta-deriv}
\end{equation}
This means that, using (\ref{eq:unimod-theta}), (\ref{eq:E4-deriv})
and (\ref{eq:Delta-deriv}), we are able to get the derivative of
the theta series of an even unimodular lattice as a function of $E_{2}$,
$E_{4}$ and $E_{6}$.

\begin{exap}
The theta series of the Gosset lattice $E_{8}$ is
\[
\Theta_{E_{8}}(\tau)=E_{4}\left(e^{-2\pi \tau}\right).
\]
Using (\ref{eq:E4-deriv}), we get
\[
{\Theta'_{E_{8}}(y)}=-\frac{2\pi}{3}\left(E_{2}\left(e^{-2\pi
\tau}\right)E_{4}\left(e^{-2\pi \tau}\right)-E_{6}\left(e^{-2\pi
\tau}\right)\right).
\]

\end{exap}
~
\begin{exap}
The theta series of the Leech lattice $\Lambda_{24}$ is
\[
\Theta_{\Lambda_{24}}(y)=E_{4}^{3}\left(e^{-2\pi
\tau}\right)-720\Delta\left(e^{-2\pi
\tau}\right)=\frac{1}{12}\left(7E_{4}^{3}\left(e^{-2\pi
\tau}\right)+5E_{6}^{2}\left(e^{-2\pi \tau}\right)\right).
\]
From this expression, and using (\ref{eq:E4-deriv}) and
(\ref{eq:E6-deriv}), we deduce the derivative with respect to $y$:
\[
{\Theta'_{\Lambda_{24}}(y)}=-\frac{\pi}{12}\left(\frac{2}{7}E_{2}\left(e^{-2\pi
\tau}\right)E_{4}^{2}\left(e^{-2\pi
\tau}\right)+5E_{2}\left(e^{-2\pi y}\right)E_{6}^{2}\left(e^{-2\pi
\tau}\right)-\frac{37}{7}E_{4}^{2}\left(e^{-2\pi
\tau}\right)E_{6}\left(e^{-2\pi \tau}\right)\right).
\]

\end{exap}

\subsubsection{The average behavior}

By the Siegel-Weyl formula, the average theta series of an even
unimodular lattice of dimension $4k$ is $E_{2k}\left(e^{-2\pi
\tau}\right)$ \cite{OSB11}. This series is a polynomial in $E_{4}$
and $E_{6}$, and so, its derivative is a polynomial in $E_{2}$,
$E_{4}$ and $E_{6}$. We give, here, a method of calculating this
polynomial.

Set
\[
G_{2k}=2\zeta\left(2k\right)E_{2k}
\]
and
\[
d_{k}=\left(2k+3\right)k!G_{2k+4}.
\]
Then, $d_{k}$ satisfies the recurrence relation,
\begin{equation}
\sum_{k=0}^{n}\dbinom{n}{k}d_{k}d_{n-k}=\frac{2n+9}{3n+6}d_{n+2}\label{eq:recurrence}
\end{equation}
with $d_{0}=3G_{4}=6\zeta(4)E_{4}$ and
$d_{1}=5G_{6}=10\zeta(6)E_{6}$. From (\ref{eq:recurrence}), we can
compute all Eisenstein series $E_{2k}$ as functions of $E_{4}$ and
$E_{6}$. Then, use (\ref{eq:E4-deriv}) and (\ref{eq:E6-deriv}) to
calculate $\nicefrac{dE_{2k}\left(e^{-2\pi \tau}\right)}{d\tau}$.

\begin{figure}[ht]
\noindent \begin{centering}
\includegraphics[width=10cm]{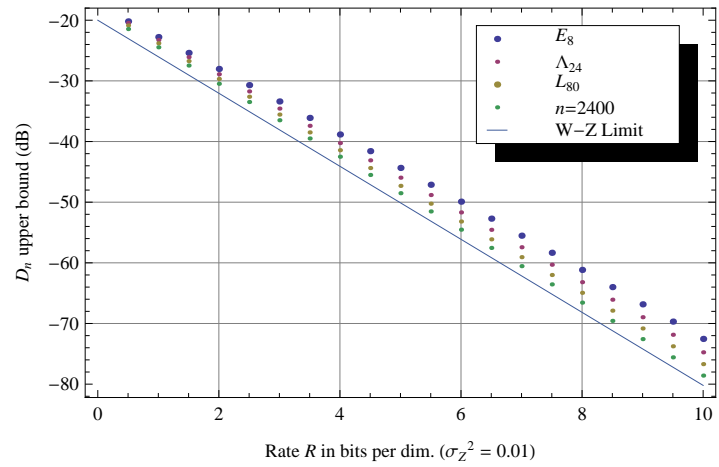}
\par\end{centering}
\caption{\label{fig:sqr} Upper bound on distortion $D_n$ versus rate
for some lattices ($n=8,24,80)$, and an average lattice ($n=2400$).
}
\end{figure}

Suppose $\sigma_{Z}^{2}=0.01$. Figure \ref{fig:sqr} shows the upper
bound on the overall distortion $D_n$ as a function of $R$ for some
lattices ($n=8,24,80$), and the average over all even unimodular
lattices of dimension $n=2400$ computed by using the derivative with
respect to $\tau$ of the Eisenstein series $E_{1200}\left(e^{-2\pi
\tau}\right)$. For the `source coding' component $D_S$, we use
Zador's upper bound on $G_n$ \cite[p.~58]{conway1998book}. It can be
seen that the upper bound on $D_n$ approaches the Wyner-Ziv limit as
$n$ increases, meaning the bound is asymptotically tight.

\section{Code Design}

In this Section, we present a variety of methods to obtain nested lattices and the associated
simulation results. In our simulation, both $X$ and $Y$ are Gaussian.
Also, $X=Y+Z$ with $Y\sim N(0,1)$ and $Z\sim N(0,0.01)$. The nested lattices in our design
may or may not be similar.

\subsection{Clean Similar Sublattices}

Recall that a clean similar sublattice means the boundary of its
 Voronoi cell does not touch any fine lattice points~\cite{conway99}. It is known that clean similar
sublattices perform better than non-clean ones~\cite{liu06}. In this
subsection we aim to find clean similar sublattices. A similar
technique was used in solving multiple-description
problem~\cite{servetto99}. In a multiple-description framework
proposed in~\cite{diggavi02}, a number of clean similar sublattice
constructions were given by extending the result in~\cite{conway99}.
We will form our own constructions to be suitably used in nested
lattices based on their constructions.

In the two-dimensional case, we introduce a complex-valued
\emph{multiplying factor} $\xi=a+bi$, $a,b \in \mathbb{Z}$, which is
multiplied to the fine lattice points to obtain the coarse lattice
points~\cite{diggavi02}. For various $\xi$, this calculation should
generate both clean and non-clean similar sublattices. The clean
similar sublattices were addressed in~\cite{diggavi02}, where the
authors proved the sublattice $\xi \mathbb{Z}^2$ is clean if and
only if the nesting ratio $a^2+b^2$ is odd, and the sublattice $\xi
A_2$ is clean if and only if $a$ and $b$ are relatively prime.

Next, we give the nested lattice construction based on
$\mathbb{Z}^4$. We use the ring of Lipschitz integer quaternions
$\mathbb{H}_0 = \{a+bi+cj+dk, a,b,c,d\in\mathbb{Z}\}$, where
$i$,$j$,$k$ are unit quaternions. Let an arbitrary point in a
four-dimensional fine lattice $\Lambda_F$ be denoted as
$\mathbf{l}_F=x+yi+zj+wk$. Then set the multiplying factor $\xi$ as
$\xi=a+bi+cj+dk$. Multiplying $\xi$ with the fine lattice points
$\mathbf{l}_F$ gives the coarse lattice points $\mathbf{l}_C$. The
calculated $\mathbf{l}_C$ is equivalent to the product shown below:
\begin{eqnarray}\label{EQ:Z4CoarseLatticePoint}
\left(
\begin{array}{cccc} a & -b & -c & -d \\
b & a & -d & c \\
c & d & a & -b \\
d & -c & b & a \\
\end{array} \right)
\cdot \left(
\begin{array}{c} x\\ y\\ z\\ w\\
\end{array} \right)
\end{eqnarray}
So if the generator matrix of the fine lattice is the identity
matrix, the first matrix above can be seen as the generator matrix
of the coarse lattice.

\begin{figure}[htbp]
  \centering
  \includegraphics{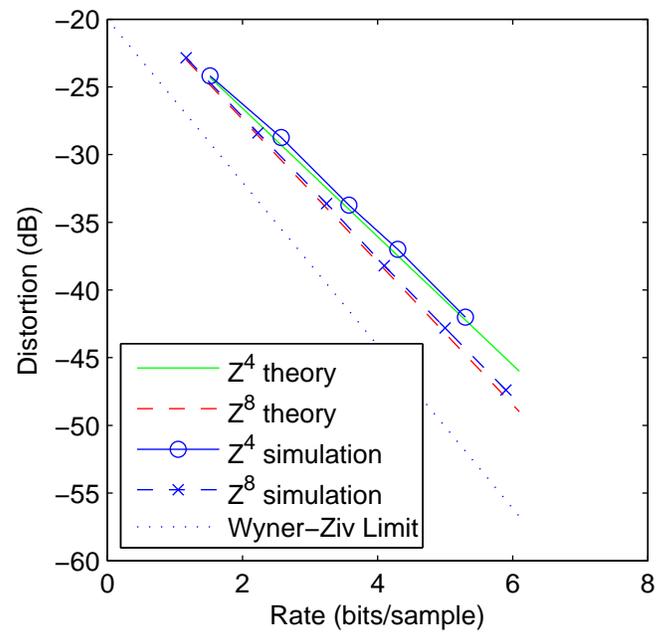}
  \caption{Distortion versus rate for clean similar sublattices of fine lattices $\mathbb{Z}^4$ and $\mathbb{Z}^8$.}
  \label{FIG:PracticeCleanSimilarSublattice}
\end{figure}

More generally, it was shown in~\cite{diggavi02} that for lattices
$\Lambda = \mathbb{Z}^{4l}, l=1,2,3 \dots$, if there exists a
geometrically similar sublattice of nesting ratio $N$, then $N$
should be of the form $m^{n/2}$ for some integer $m$, where $n=4l$
is the dimension. From~\cite{diggavi02}, the nesting ratio $N$ must
be odd to make sure the similar sublattices are clean. From our
simulation, the nesting ratios which are odd perfect squares indeed
give better performance than others. The simulation results are
shown in Fig.~\ref{FIG:PracticeCleanSimilarSublattice}.

\subsection{Obtaining Similar Sublattices by Scaling and Rotation}

In this subsection, we present a more general method to obtain
sublattices by scaling and rotation of the fine lattice. The
obtained sublattices are not necessarily clean. We set a parameter,
$\beta$, as the \emph{expanding factor}. The way to generate the
similar coarse lattice is as follows:

\begin{itemize}
\item First
expand the fine lattice by multiplying the expanding factor $\beta$,
such that the lattice points with the smallest norm greater than
zero in the fine lattice become the outermost points in the basic
Voronoi cell of the coarse lattice. When $\beta$ is an integer,
there is always a less (or equal) number of coarse lattice points on
each shell (formed by points of equal norm) of the coarse lattice
than in the fine lattice. This requirement can be fulfilled by using
integer expanding factors.
\item For non-integer expanding factors, some expanded lattice points may not be a subset of the fine
lattice. This case does not only happen to only one point, but also
to a group of points with the same norm. The angle between these
points and their fine lattice counterparts can be detected and
calculated. Due to symmetry, these angles for points with the same
norm are the same. So we can rotate the expanded lattice points by
this angle, to match the positions of those fine lattice points.
Also, the rotation for all the norms should be the same since the
expanded lattice is similar to the fine one; so the rotated coarse
lattice remains the similar shape as the original fine lattice.
\end{itemize}

The values of $\beta$ giving rise to sublattices are those norms of
the fine lattices. So $\beta$ is not necessarily an integer. By
exploiting tables of theta series in~\cite{conway1998book}, we can
find a sublattice.

\begin{figure}[htbp]
\centering \subfigure[Before
rotation]{\label{fig:subfig1}\includegraphics[width=6cm]{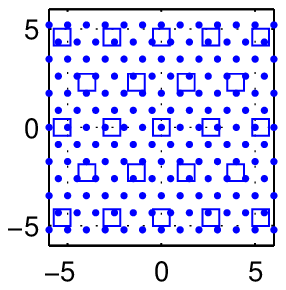}}
\subfigure[After
rotation]{\label{fig:subfig2}\includegraphics[width=6cm]{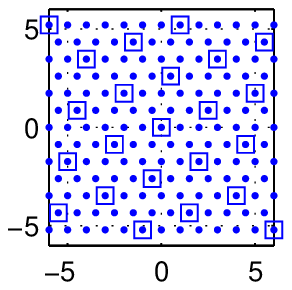}}
 \caption[Optional caption for list of figures]{Example of two-dimensional coarse lattice: $\beta=\sqrt{7}$.
 $\square$ coarse lattice points, $\bullet$ fine lattice points.}
\label{fig:peqs7}
\end{figure}

\begin{exap}

We give an example in the two-dimensional case for both
integer and non-integer expanding factors. Let the fine lattice be
a hexagonal lattice generated by the following generator matrix:
\begin{eqnarray}\label{eq:2DLatGenMat}
\mathbf{M}_F=\left(
\begin{array}{ccc}
1 & -\frac{1}{2} \\
0 & \frac{\sqrt 3}{2}
\end{array} \right)
\end{eqnarray}

Then the fine lattice is expanded (and rotated when necessary) to
form the coarse lattice $\mathbf{M}_C=\beta \mathbf{M}_F$. For fine
and coarse lattices with integer expanding factors (e.g. $\beta=3$
and $\beta=4$), rotation is not needed. When $\beta=\sqrt{7}$,
rotation is needed to make the coarse lattice points form a subset
of the fine lattice, as shown in Fig.~\ref{fig:peqs7}.

\end{exap}

Using this method, we now give the
simulation results of the rate-distortion performance for nested
lattice Wyner-Ziv coding for $D_n$, $E_8$ and $\Lambda_{24}$.

\begin{figure}[htbp]
  \centering
  \includegraphics{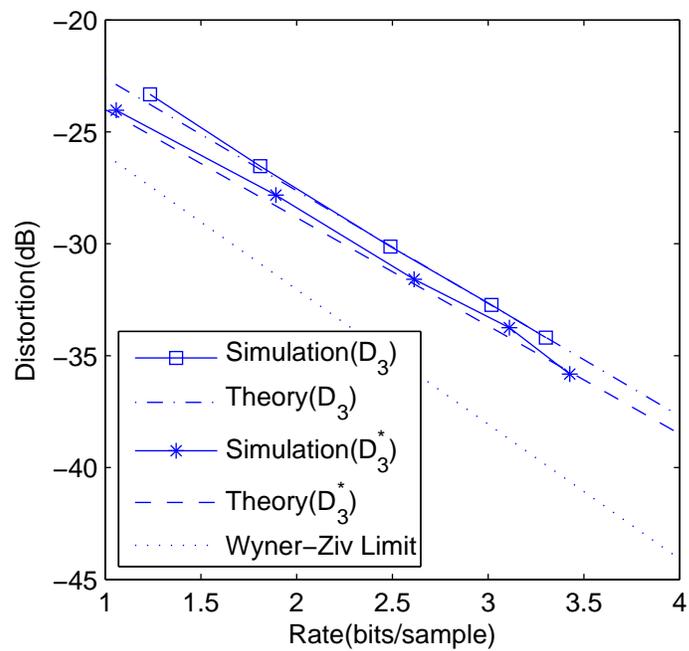}
  \caption{Distortion versus rate for fine lattices $D_3$ and $D_3^\ast$.}
  \label{fig:DvsR3D}
\end{figure}

In the three-dimensional case, we implement the scheme using both
$D_3$ and $D_3^\ast$. The rate-distortion performance is shown in
Fig.~\ref{fig:DvsR3D}. Compared to the one- and two-dimensional
case, the three-dimensional scheme gives less distortion. Also
notice NLC using $D_3^\ast$ is better than the one using $D_3$.

The simulation results using $D_4$ and $E_8$ are similar, and are
both better than the three-dimensional case. At rate $1.5$ bits per
sample, their distortion gaps to the Wyner-Ziv limit are $2.52$ dB
and $1.07$ dB, respectively.

\begin{figure}[htbp]
  \centering
  \includegraphics{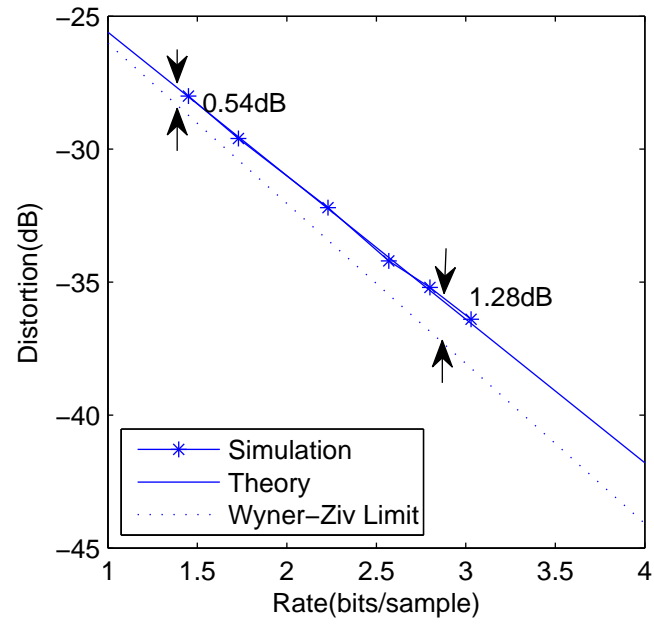}
  \caption{Distortion versus rate for fine lattice $\Lambda_{24}$.}
  \label{fig:DvsR24D}
\end{figure}

Moreover, for $\Lambda_{24}$, our scheme gives distortion
performance closer to the Wyner-Ziv limit than the SWC-NQ scheme
proposed in~\cite{liu06}, especially at low rates. Results
in~\cite{liu06} show a gap of distortion $1.53$ dB from the
Wyner-Ziv limit while our gap is less than $1.28$ dB at rate less
than three bits per sample for $\Lambda_{24}$. See
Fig.~\ref{fig:DvsR24D}.

One can also see that the ``increasing gap" between the
rate-distortion curve of NLC and the Wyner-Ziv limit~\cite{liu06}
indeed exists in concrete implementation. Nonetheless, from
Fig.~\ref{fig:DvsR24D}, this widening gap as rate increases can be
handled by increasing the dimension. In the meantime, the rate can
not be too high in sensor network applications. So the gap is
acceptable with $8$- or $24$-dimensional lattices. By increasing the
dimension of NLC as the rate increases, a constant gap from the
Wyner-Ziv limit can be maintained.

\subsection{Ensemble of Random Nested Lattices}

Code designs above are restricted to those proposed
in~\cite{conway1998book}. Although they have acceptable performance
and low complexity, it is better to have a scheme existing in any
dimensions. Hence, we use the ensemble of good nested lattice codes
proposed in~\cite{erez04} based on the concept of random lattices.
The random lattice ensemble in~\cite{erez04} can be generated as
follows.

\begin{itemize}
\item Take $p$ to be prime.

\item Define a $k\times n$ generator matrix $M$, where $M_{i,j}$ is uniformly distributed on $(0,\dots,p-1),i=1,\dots,k;j=1,\dots,n$.

\item Apply Construction A in~\cite{erez04} to obtain the lattice
$\Lambda_F^{\prime}$.
\end{itemize}

The $n$-dimensional cubic lattice $\mathbb{Z}^{n}$ can be viewed as
a sublattice of the random lattice
$\Lambda_F^{\prime}$~\cite{erez04}. Hence, for any dimension,
various nested lattices can be obtained by simply applying different
linear transformations $\mathbf{G}$ to both $\mathbb{Z}^{n}$ and
$\Lambda_F^{\prime}$. Obviously, the resultant sublattice is not
necessarily similar. The nest ratio $N=p^{n-k}$.

To make the nested ensemble good for the Wyner-Ziv problem, the fine
lattice should be good for source coding and the coarse lattice good
for channel coding. This is proved in~\cite{krithivasan09} by
extending the results in~\cite{erez04}. Also, both the coarse
lattice $\mathbf{G} \mathbb{Z}^{n}$ and the fine lattice $\mathbf{G}
\Lambda_F^{\prime}$ should be good for quantization. To make
$\mathbf{G} \mathbb{Z}^{n}$ good for quantization, $\mathbf{G}$
should be the generator matrix of a good quantizing
lattice~\cite{conway1998book}, e.g., the hexagonal lattice in
dimension two and $E_8$ in dimension eight.

However, such lattices are not easy to decode. We use the sphere
decoding algorithm described in~\cite{SchnorrEuchner} for the
quantization to the random lattices, whose speed is tolerable if $n
\leq 24$.

\begin{figure}[htbp]
  \centering
  \includegraphics{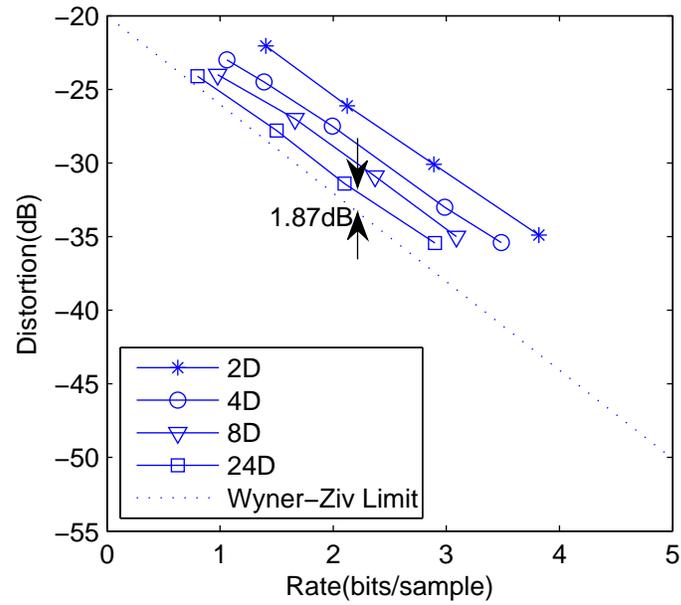}
  \caption{Distortion versus rate for the ensemble of random nested lattices.}
  \label{FIG:PracticeRandomNestedEnsembles}
\end{figure}

Examples for dimensions $2$, $4$, $8$ and $24$ are given in
Fig.~\ref{FIG:PracticeRandomNestedEnsembles}. The rate-distortion
performance approaches the Wyner-Ziv limit as dimension increases
and the gap is only $1.87$ dB within the limit for dimension
twenty-four. The performance is very close to explicit
lattices like the Leech lattice.

Since the elements of $\mathbf{M}$ are uniformly distributed over $\mathbb{Z}_{p}$
for large $p$ and since the decoding is not easy, this ensemble is of theoretic value only.

\section{Conclusions}

In this paper, we have investigated the rate-distortion function and
practical code design of Wyner-Ziv coding based on multi-dimensional
nested lattices. Under the high-resolution assumption, an accurate
calculation was developed, and an upper bound expressed in terms of
the derivative of the theta series was derived. These results can be
used to judge the performance and serve as a practical guide for
choosing good lattices for Wyner-Ziv coding. Several practical code
designs were presented using multi-dimensional NLC. High-dimensional
schemes gave performance close to the Wyner-Ziv limit. The
performance is even better than the SWC-NQ scheme~\cite{liu06}.
Compared to ~\cite{liu06}, our scheme does not require extra
Slepian-Wolf coding based on powerful error correction codes,
thereby enjoying lower complexity. Hence, the proposed scheme may be
attractive for applications in sensor networks where simple coding
schemes are needed.

This work leaves some open problems. Firstly, there may be room to
improve the upper bound. Secondly, the derivative of the theta
series arising from the upper bound, which is to be maximized, is a
new problem for lattice researchers. Last but not the least, a more
systematic approach to low-complexity code design is to be pursued.

\bibliographystyle{IEEEtran}
\bibliography{SuGaoRef}

\begin{thebibliography}{10}
\providecommand{\url}[1]{#1}
\csname url@samestyle\endcsname
\providecommand{\newblock}{\relax}
\providecommand{\bibinfo}[2]{#2}
\providecommand{\BIBentrySTDinterwordspacing}{\spaceskip=0pt\relax}
\providecommand{\BIBentryALTinterwordstretchfactor}{4}
\providecommand{\BIBentryALTinterwordspacing}{\spaceskip=\fontdimen2\font plus
\BIBentryALTinterwordstretchfactor\fontdimen3\font minus
  \fontdimen4\font\relax}
\providecommand{\BIBforeignlanguage}[2]{{%
\expandafter\ifx\csname l@#1\endcsname\relax
\typeout{** WARNING: IEEEtran.bst: No hyphenation pattern has been}%
\typeout{** loaded for the language `#1'. Using the pattern for}%
\typeout{** the default language instead.}%
\else
\language=\csname l@#1\endcsname
\fi
#2}}
\providecommand{\BIBdecl}{\relax}
\BIBdecl

\bibitem{xiong04}
Z.~Xiong, A.~D. Liveris, and S.~Cheng, ``Distributed source coding for sensor
  networks,'' \emph{IEEE Signal Processing Magazine}, vol.~21, no.~5, pp.
  80--94, 2004.

\bibitem{slepian73}
D.~Slepian and J.~Wolf, ``Noiseless coding of correlated information sources,''
  \emph{IEEE Trans. Inform. Theory}, vol.~19, no.~4, pp. 471--480, 1973.

\bibitem{wyner76}
A.~Wyner and J.~Ziv, ``The rate-distortion function for source coding with side
  information at the decoder,'' \emph{IEEE Trans. Inform. Theory}, vol.~22,
  no.~1, pp. 1--10, 1976.

\bibitem{Zamir98}
R.~Zamir and S.~Shamai, ``Nested linear/lattice codes for {Wyner-Ziv}
  encoding,'' \emph{Proc. Inform. Theory Workshop, Killarney, Ireland}, pp.
  92--93, June 1998.

\bibitem{zamir02}
R.~Zamir, S.~Shamai, and U.~Erez, ``Nested linear/lattice codes for structured
  multiterminal binning,'' \emph{IEEE Trans. Inform. Theory}, vol.~48, no.~6,
  pp. 1250--1276, 2002.

\bibitem{servetto00}
S.~D. Servetto, ``Lattice quantization with side information,'' \emph{Proc.
  IEEE Data Compression Conference (DCC)}, 2000.

\bibitem{liu06}
Z.~Liu, S.~Cheng, A.~D. Liveris, and Z.~Xiong, ``Slepian-{W}olf coded nested
  lattice quantization for {W}yner-{Z}iv coding: High-rate performance analysis
  and code design,'' \emph{IEEE Trans. Inform. Theory}, vol.~52, pp.
  4358--4379, 2006.

\bibitem{conway99}
J.~H. Conway, E.~M. Rains, and N.~J.~A. Sloane, ``On the existence of similar
  sublattices,'' \emph{Canad. J. Math.}, vol.~51, pp. 1300--1306, 1999.

\bibitem{erez04}
U.~Erez and R.~Zamir, ``Achieving (1/2) log(1+{SNR}) on the {AWGN} channel with
  lattice encoding and decoding,'' \emph{IEEE Trans. Inform. Theory}, vol.~50,
  no.~10, pp. 2293--2314, 2004.

\bibitem{fiore10}
P.~D. Fiore, ``Scale-recursive lattice-based multiple-access symbol
  constellations,'' \emph{IEEE Trans. Inform. Theory}, vol.~56, no.~1, pp.
  211--223, 2010.

\bibitem{sartipi08}
M.~Sartipi and F.~Fekri, ``Distributed source coding using short to moderate
  length rate-compatible {L}{D}{P}{C} codes: The entire {S}lepian-{W}olf rate
  region,'' \emph{IEEE Trans. Commun.}, vol.~56, no.~3, pp. 400--411, 2008.

\bibitem{yang09}
Y.~Yang, S.~Cheung, Z.~Xiong, and W.~Zhao, ``Wyner-{Z}iv coding based on {TCQ}
  and {LDPC} codes,'' \emph{IEEE Trans. Commun.}, vol.~57, no.~2, pp. 376--387,
  2009.

\bibitem{conway1998book}
J.~H. Conway and N.~J.~A. Sloane, \emph{Sphere Packings, Lattices and
  Groups}.\hskip 1em plus 0.5em minus 0.4em\relax Springer-Verlag, 1998.

\bibitem{OSB11}
\BIBentryALTinterwordspacing
F.~Oggier, P.~Sol{\'e}, and J.-C. Belfiore, ``Lattice codes for the wiretap
  {G}aussian channel: Construction and analysis,'' Mar. 2011. [Online].
  Available: \url{http://arxiv.org/abs/1103.4086}
\BIBentrySTDinterwordspacing

\bibitem{Ebeling}
W.~Ebeling, \emph{Lattices and Codes}.\hskip 1em plus 0.5em minus 0.4em\relax
  Vieweg, 1994.

\bibitem{Ramanujan}
S.~Ramanujan, ``On certain arithmetical functions,'' \emph{Trans. Cambridge
  Philos. Soc.}, vol.~22, p. 159\textendash{}184, 1916.

\bibitem{servetto99}
S.~D. Servetto, V.~A. Vaishampayan, and N.~J.~A. Sloane, ``Multiple description
  lattice vector quantization,'' \emph{Data Compression Conference}, pp.
  13--22, 1999.

\bibitem{diggavi02}
S.~N. Diggavi, N.~J.~A. Sloane, and V.~A. Vaishampayan, ``{A}symmetric multiple
  description lattice vector quantizers,'' \emph{IEEE Trans. Inform. Theory},
  vol.~48, no.~1, pp. 174--191, 2002.

\bibitem{krithivasan09}
D.~Krithivasan and S.~S. Pradhan, ``Lattices for distributed source coding:
  Jointly {G}aussian sources and reconstruction of a linear function,''
  \emph{IEEE Trans. Inform. Theory}, vol.~55, no.~12, pp. 5628--5651, 2009.

\bibitem{SchnorrEuchner}
C.~P. Schnorr and M.~Euchner, ``Lattice basis reduction: Improved practical
  algorithms and solving subset sum problems,'' \emph{Math. Program.}, vol.~66,
  pp. 181--191, 1994.

\end{thebibliography}


\newpage

\end{document}